\documentclass[default,iicol]{sn-jnl}% Default with double column layout

\pdfoutput=1
\hypersetup{pdfauthor={Hossein Jafari},pdftitle={Swin-Tempo: Temporal-Aware Lung Nodule Detection in CT Scans as Video Sequences Using Swin Transformer-Enhanced UNet}}

\usepackage{graphicx}%
\usepackage{multirow}%
\usepackage{makecell}
\usepackage{amsthm}
\usepackage{dcolumn}

\usepackage{subcaption}

\usepackage{cellspace}
\usepackage{arydshln}
\usepackage{natbib}
\usepackage{amsmath,amssymb,amsfonts}%
\usepackage{amsthm}%
\usepackage{mathrsfs}%
\usepackage[title]{appendix}%
\usepackage{graphicx}
\usepackage{xcolor}%
\usepackage{textcomp}%
\usepackage{manyfoot}%
\usepackage{placeins}
\usepackage{booktabs}%
\usepackage{algpseudocode}%
\usepackage{listings}%
\usepackage{orcidlink}
\usepackage[linesnumbered,ruled,vlined]{algorithm2e}

\theoremstyle{thmstyleone}%
\theoremstyle{thmstyletwo}%

\theoremstyle{thmstylethree}%

\begin{document}

\title[Swin-Tempo: Temporal-Aware Lung Nodule Detection in CT Scans as Video Sequences Using Swin Transformer-Enhanced UNet]{Swin-Tempo: Temporal-Aware Lung Nodule Detection in CT Scans as Video Sequences Using Swin Transformer-Enhanced UNet}


%\author{Hossein Jafari\textsuperscript{\orcidlink{0009-0000-7345-2233}}}

%\author{Karim Faez\textsuperscript{\orcidlink{0000-0002-1159-4866}}}
%\author{Hamidreza Amindavar\textsuperscript{\orcidlink{0000-0002-0954-0674}}}

\author[]{\fnm{Hossein} \sur{Jafari}\textsuperscript{\orcidlink{0009-0000-7345-2233}} } 

\author*[]{\fnm{Karim} \sur{Faez}\textsuperscript{\orcidlink{0000-0002-1159-4866}} } \email{karimfaez@aut.ac.ir}

\author[]{\fnm{Hamidreza} \sur{Amindavar}\textsuperscript{\orcidlink{0000-0002-0954-0674}} }

\affil[]{\orgdiv{Department of Electrical Engineering}, \orgname{Amirkabir University of Technology}, \city{Tehran}, \country{Iran}}

\abstract{Lung cancer is highly lethal, emphasizing the critical need for early detection. However, identifying lung nodules poses significant challenges for radiologists, who rely heavily on their expertise for accurate diagnosis. To address this issue, computer-aided diagnosis (CAD) systems based on machine learning techniques have emerged to assist doctors in identifying lung nodules from computed tomography (CT) scans. Unfortunately, existing networks in this domain often suffer from computational complexity, leading to high rates of false negatives and false positives, limiting their effectiveness. To address these challenges, we present an innovative model that harnesses the strengths of both convolutional neural networks and vision transformers. Inspired by object detection in videos, we treat each 3D CT image as a video, individual slices as frames, and lung nodules as objects, enabling a time-series application. The primary objective of our work is to overcome hardware limitations during model training, allowing for efficient processing of 2D data while utilizing inter-slice information for accurate identification based on 3D image context.
We validated the proposed network by applying a 10-fold cross-validation technique to the publicly available Lung Nodule Analysis 2016 dataset. Our proposed architecture achieves an average sensitivity criterion of \textbf{97.84\%} and a competition performance metrics (CPM) of \textbf{96.0\%} with few parameters. Comparative analysis with state-of-the-art advancements in lung nodule identification demonstrates the significant accuracy achieved by our proposed model.}


\maketitle

\newpage
\section{Introduction}\label{sec1}

Among the leading causes of cancer-related mortality worldwide, lung cancer remains a significant concern for global health. A study by the American Cancer Society concluded that out of approximately 600,000 deaths, more than 21 percent were due to lung cancer \cite{Siegel2022}. This highlights the profound impact of lung cancer nodules on public health, underscoring the importance of implementing effective screening, early detection, and treatment interventions.

The early detection of lung nodules can significantly enhance treatment outcomes and improve patient survival rates \cite{huang2023artificial}. To address this, CT scans have emerged as the primary and widely adopted approach for detecting pulmonary nodules \cite{naik2021lung,xu2020deepln,rocha2020conventional,mei2021sanet}. However, due to the extensive nature of CT data, with hundreds of slices per scan, radiologists face a substantial challenge in thoroughly examining each slice to identify potential nodules \cite{zhou2022cascaded}. CAD systems leveraging artificial intelligence (AI) have been developed to address this demanding task, identifying meaningful patterns and facilitating precise nodule detection \cite{li2022deep, ye2020pulmonary, sharif2020comprehensive}. These sophisticated algorithms analyze CT data and facilitate the detection and classification of pulmonary nodules, thereby enhancing the efficiency and accuracy of the diagnostic process. However, this endeavor is complex, given the diverse range of challenges posed by variations in nodule size, shape, density, and anatomical context, as well as the presence of tissue structures that closely approximate the appearance of nodules, all of which complicate the task of automated detection \cite{luo2022scpm}. Recent advances in convolutional neural networks (CNNs) and transformer-based models have emerged as cutting-edge methods for the automatic detection and segmentation of pulmonary nodules, demonstrating substantial success in this domain \cite{mkindu2023lung, xu2023sgda, wu2023medical, abid2021multi, agnes2022two, su2021lung}.  Moreover, researchers have focused on developing multi-modal strategies that integrate diverse imaging techniques, further augmenting the precision of pulmonary nodule detection and enhancing overall diagnostic capabilities \cite{agnes2022two}.

\par In the context of processing CT scans, CAD systems face inherent limitations, with one key factor being restricted graphical processing units (GPU) memory \cite{brogan2021deep}. These limitations prompt the adoption of two distinct strategies: slice-based processing and volume-based processing. For slice-based processing, the initial approach involves analyzing each slice within the CT scan. 2D models, being able to view the entire slice at once, possess the advantage of learning the structural intricacies of the lung and capturing dependencies between different regions. This capability enables them to comprehensively understand the lung's structure, enhancing their ability to detect abnormalities accurately \cite{xie2019automated, zheng2019automatic}. On the other hand, volume-based processing, typically employed by 3D models, involves analyzing the entire 3D CT volume. Due to the need for cropping the image to overcome the limitations of GPU in terms of memory and processing \cite{nasrullah2019automated, riquelme2020deep, zhang2022pulmonary}, 3D models may not thoroughly learn the lung's intricate structure. However, they compensate for this limitation by accessing multiple slices, allowing them to view objects from different perspectives. This unique capability enables 3D models to extract features from various slices, facilitating the representation of robust and reliable candidate regions \cite{zhao2023attentive,gong2019automated,mittapalli2021multiscale}.

\par Additionally, in the domain of nodule detection and segmentation tasks, CNN-based models have gained prominence due to their efficient parameter-sharing mechanism. By sharing the same weights across the entire image, CNNs significantly reduce the number of learnable parameters, resulting in efficient training and inference and making them scalable to large datasets \cite{alzubaidi2021review}. Their local feature extraction has limitations in capturing long-range dependencies and the global context \cite{zhao2023swingan}, whereas transformer-based models offer a notable advantage due to their inherent parallelization capability. On the other hand, transformer-based models present a unique advantage with their inherent parallelizable nature. Leveraging self-attention mechanisms \cite{vaswani2017attention}, transformers can simultaneously model interactions between all pixels, facilitating efficient parallel processing during training and inference. This parallelizability empowers transformers to handle complex spatial relationships and capture long-range dependencies effectively. Nonetheless, transformers may require more significant memory and computational resources, especially in dense pixel-level segmentation tasks. Unlike recurrent neural networks (RNN) that process sequence elements sequentially and have restricted access to contextual information \cite{liu2019bidirectional}, transformer-based models can consider the entire sequence simultaneously, enabling them to capture long-range dependencies and excel in scenarios with objects distributed across images at varying scales and locations \cite{Swin-basic2}. Combined with multi-head attention for parallel processing, this sets transformers apart from other neural networks. The scalability of transformers sets them apart from CNN-based models, making them suitable for complex models and extensive datasets \cite{mkindu2023lung}.

\par To address the mentioned problems, we introduce a new end-to-end framework, Swin-Tempo, for online pulmonary nodule detection from CT images. This study offers the following key contributions:

\begin{enumerate}
\item{Our framework leverages video-based processing, treating each CT image as a video and each slice of the CT as a frame of that video, enabling slice-by-slice nodule detection and tracking. By adopting this video-based approach, we harness the advantages of both 2D and 3D models to capture temporal dependencies effectively.
}

\item{Our proposal incorporates a hybrid CNN-transformer model, enabling us to combine the strengths of CNN and transformer-based models in single-frame processing. This integration facilitates efficient feature extraction and global context modeling, contributing to enhanced accuracy and real-time capabilities in nodule detection.}

\end{enumerate}

In the subsequent sections, we provide a comprehensive overview of our model's design, methodology, experimental results, and comparative evaluations, demonstrating its superior performance and potential impact on medical imaging.

\section{Methods}\label{sec2}
\par In this work, we adopt a novel approach that considers a CT scan as a video sequence, where each CT slice corresponds to a frame of the video. This perspective enables us to perform online object detection within the video. It opens up exciting possibilities for applying advanced techniques commonly utilized in video analysis, such as temporal convolutional networks and recurrent neural networks. These methods strategically leverage temporal information to represent dynamic patterns and enable motion-aware nodule detections in CT slices.
\par
\begin{figure}[]
\centering
\includegraphics[height=0.31\textheight]{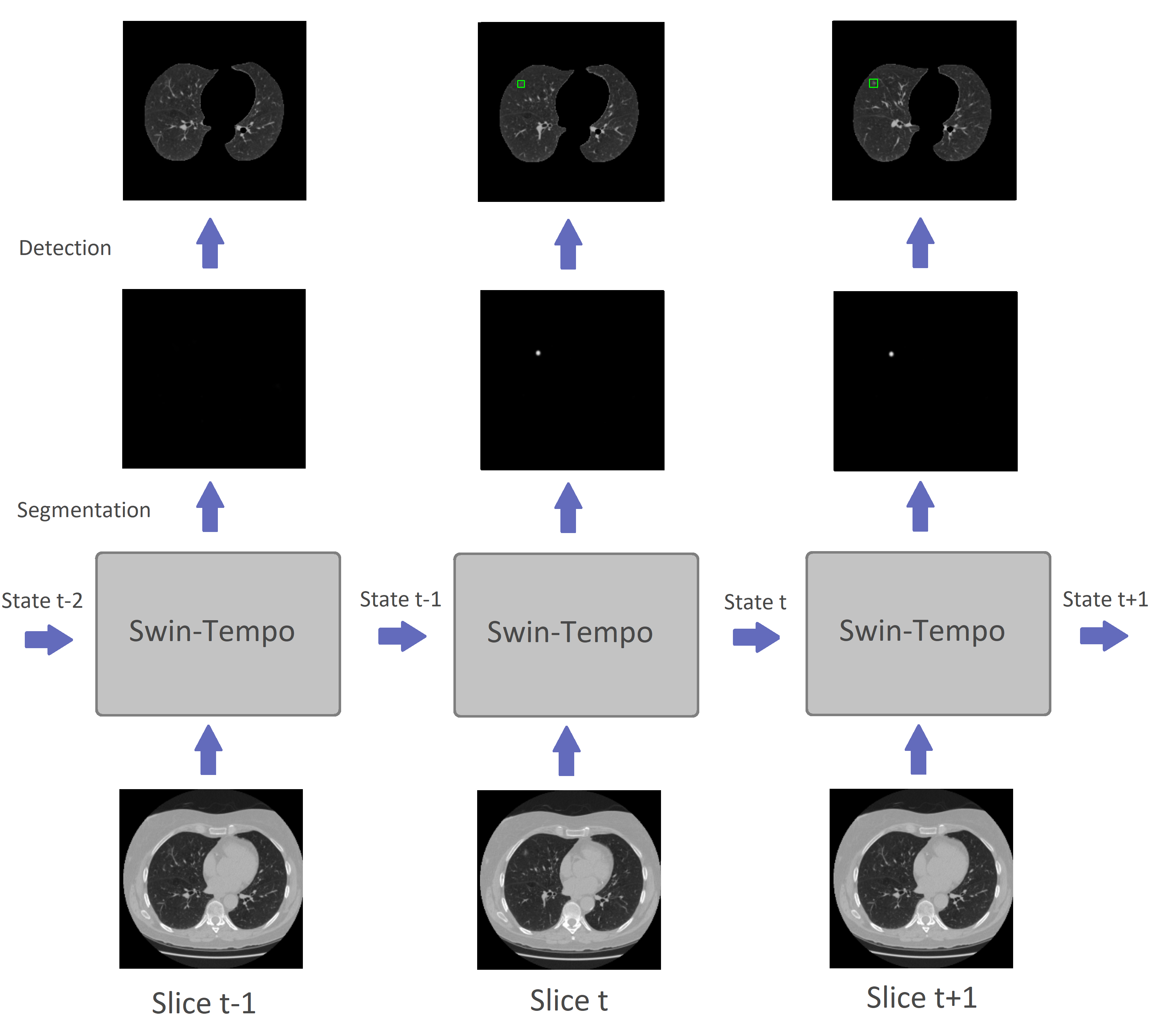}
\caption{Swin-Tempo model architecture. An illustration showcasing the sequential processing of three CT slices and its use of temporal information to improve pulmonary nodule detection.}
\label{tempo}
\end{figure}

 This advancement brings several benefits to the analysis process, including improved tracking, a better understanding of complex actions, handling varying frame rates, reducing redundancy, adapting to irregular time intervals between frames, and effectively addressing the constraints imposed by limited GPU memory. Consequently, our proposed approach facilitates real-time and efficient analysis. Our method introduces new opportunities for enhancing pulmonary nodule detection in medical imaging by combining video-based analysis with hybrid models. Figure \ref{tempo} illustrates how our model processes a CT image and generates outputs.

Our method intelligently integrates the strengths of both 2D and 3D models. Inspired by 2D models, we implement a slice-level processing strategy, allowing us to handle individual CT slices efficiently and comprehensively capture essential lung structures and dependencies. At the same time, we incorporate critical elements of 3D models by accessing multiple slices, enabling us to extract features from diverse perspectives and ensure a robust representation of nodules. This pragmatic and versatile approach achieves high accuracy and efficiency in detection, leading to timely diagnoses and improved clinical outcomes.

\par In Section \ref{subsec:swin-tempo-key-modules}, we will start by explaining the three key modules that make up the foundation of the proposed architecture, called Swin-Tempo. These modules are essential components that each bring unique strengths to our proposed approach. Firstly, we will discuss the Swin Transformer, a cutting-edge deep-learning architecture that captures high-level spatial features from images. Secondly, we will explore the UNet \cite{UNet-Basic}, a versatile and widely-used architecture renowned for its ability to capture both local and global context understanding in medical image analysis. Lastly, we will elaborate on the gated recurrent unit (GRU), a neural network aware of time and excels in modeling sequential data and capturing temporal dependencies.

\par Once these fundamental modules are thoroughly explained in Section \ref{subsec:swin-tempo-key-modules}, we will transition to the subsequent subsections to unveil our three distinct approaches: Integrating Swin Transformer with UNet's encoder, Integrating GRU with UNet's decoder. Each method innovatively integrates the abovementioned modules, harnessing their strengths to enhance pulmonary nodule detection from CT images. We will describe each methodology, network architecture, and unique contributions that set them apart.

\subsection{Swin-Tempo Architecture: Key Modules}
\label{subsec:swin-tempo-key-modules}

\begin{figure*}[]
	\centering
	\includegraphics[height=0.93\textheight]{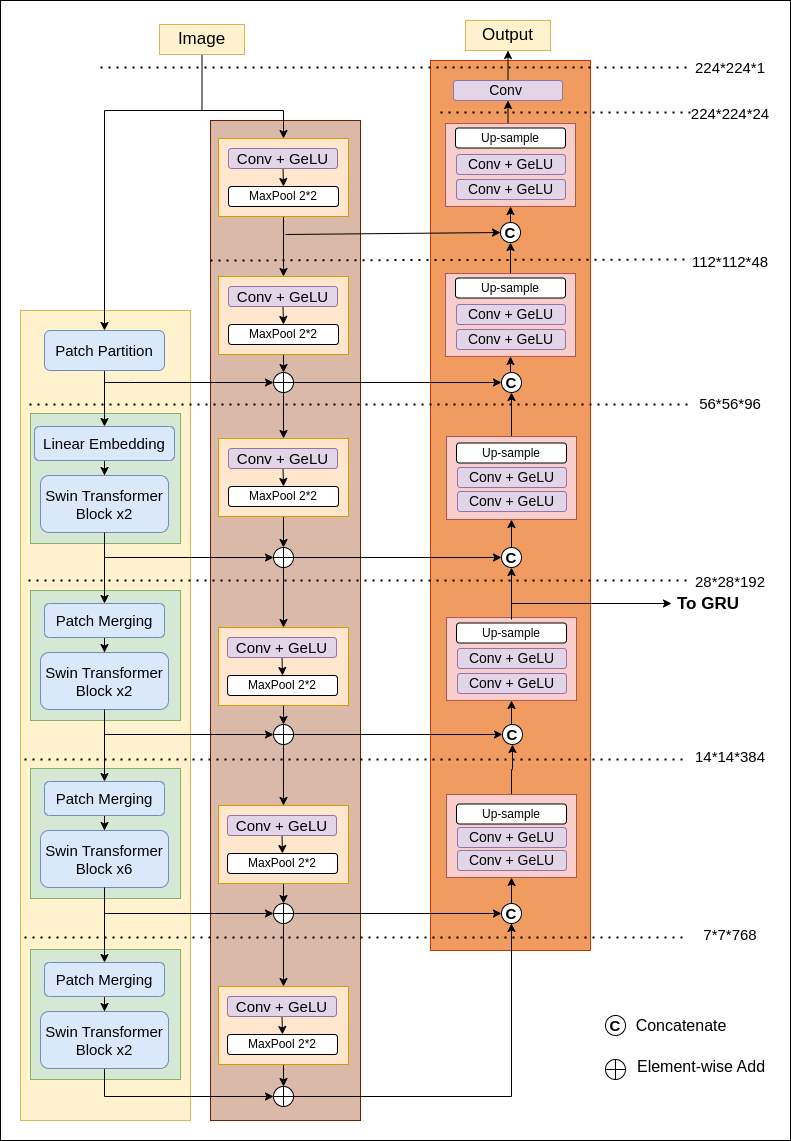}
	\caption{ Swin-Tempo Structure.}
	\label{final}
\end{figure*}

The Swin-Tempo architecture captures low-level and high-level features from the input data and reconstructs the output with refined context information. To achieve this, it incorporates the UNet encoder and decoder with Swin Transformer and GRU, respectively. This integration helps the model better understand the context and structures in CT scans, contributing to its exceptional performance in nodule detection tasks. The overview of our proposed model architecture is shown in Figure \ref{final}.

In the following subsections, we will thoroughly explain the Swin-Tempo structure, highlighting the three essential components that make up its core. These modules work together to improve the efficiency of our method for identifying pulmonary nodules from CT scans.

\subsubsection{Swin Transformer}

The Swin Transformer is a remarkable deep-learning architecture with a pronounced ability to extract high-level spatial features from images \cite{liu2021swin}. This innovative model has proven useful in various computer vision applications, such as object detection, image segmentation, and image classification, primarily owing to its effective self-attention mechanisms \cite{Swin-basic, Swin-Eg3}. The model's ingenious approach involves efficiently processing 2D images through non-overlapping patch division and multiple stages of transformer layers, consistently outperforming traditional convolutional neural networks on benchmark datasets \cite{liu2021swin}.

In the Swin Transformer architecture, the initial step entails the meticulous division of each input RGB image into non-overlapping patches, with each patch represented by a 48-dimensional feature vector obtained by concatenating the raw pixel RGB values. This patch-based processing allows the Swin Transformer to capture fine-grained information from the image and addresses position insensitivity during self-attention computations \cite{dosovitskiy2020image}. In order to create a hierarchical representation of the input data, patch merging layers are incorporated as the network progresses through various stages. This results in feature maps with varying dimensions to capture features at different scales effectively \cite{Swin-Eg1, Swin-Eg2}.

Critical to the Swin Transformer's efficiency is the process of patch merging \cite{liu2021swin}. Patch merging layers are strategically placed at each stage of the model to reduce the number of tokens and achieve resolution downsampling. For instance, the first patch merging layer concatenates neighboring patch features and employs a linear layer to produce a 2C-dimensional output, achieving a 2x downsampling resolution. These layers are pivotal in streamlining the model's computations \cite{Swin-Eg1, Swin-Eg2}. Within the Swin Transformer block, a series of mathematical operations are performed:
\begin{equation}
\begin{aligned}
    \text{Step 1:} \quad & \hat{z}_l = \text{W-MSA}(\text{LN}(z_{l-1})) + z_{l-1},  \\
    \text{Step 2:} \quad & z_l = \text{MLP}(\text{LN}(\hat{z}_l)) + \hat{z}_l, \\
    \text{Step 3:} \quad & \hat{z}_{l+1} = \text{SW-MSA}(\text{LN}(z_l)) + z_l,  \\
    \text{Step 4:} \quad & z_{l+1} = \text{MLP}(\text{LN}(\hat{z}_{l+1})) + \hat{z}_{l+1}, 
\end{aligned}
\end{equation}
Here, $\hat{z}_l$ and $z_l$ represent intermediate and final hidden states, respectively. $l = 1, 2, ..., L$ represents the layer number, and $L$ is the total number of layers. W-MSA stands for Window-based Multi-head Self-Attention, LN denotes Layer Normalization, and MLP represents Multi-Layer Perceptron.

\subsubsection{UNet}
\par UNet is a widely adopted CNN-based architecture extensively used in medical image analysis \cite{UNet-Basic, UNet++}. It comprises two essential components: the contracting and expanding paths.
\par In the contracting path, the network reduces the input image's spatial dimensions by half through 2*2 max-pooling, facilitating context capture and expanding the receptive field. Subsequently, down-sampled feature maps undergo further refinement via a 3*3 convolutional layer, effectively extracting vital features from localized regions. Conversely, the expanding path focuses on precise localization and segmentation map reconstruction. It employs up-sampling and 3*3 transposed convolutions to increase the spatial dimensions of the feature maps. This results in a finely detailed segmentation output that aligns seamlessly with the original input image \cite{UNetplus}.
\par A vital strength of the UNet lies in its skip connections, which connect the contracting and expanding paths. These connections significantly enhance the model's capacity to capture intricate details and contextual information, contributing substantially to its superior performance in generating precise and accurate segmentations. This restorative process allows the model to generate dense pixel-wise predictions for each CT slice, accurately delineating potential nodule boundaries and localizing their positions within the CT scan \cite{UNet-CTScan}. By effectively retaining and fusing crucial original image information with decoded features, the UNet demonstrates its remarkable effectiveness and reliability across various applications, particularly in medical image analysis.

\subsubsection{Gated Recurrent Unit}
\par GRU, a type of recurrent neural network architecture, has gained widespread popularity for its effectiveness in modeling sequential data. It has become a prominent choice for a variety of sequential tasks, including natural language processing \cite{GRU-Basic, GRU-Eg1}, speech recognition \cite{GRU-Eg2}, and time series analysis \cite{GRU-EG3}, among others. GRU prove particularly valuable when dealing with sequential data characterized by temporal dependencies or patterns.

%###########Integrating Swin Transformer with UNet's encoder################
\subsection{Integrating Swin Transformer with UNet's Encoder}
\par In our study, we developed a novel approach by combining two neural networks, Swin Transformer and UNet, to process individual slices of CT scans. This integration resulted in a unique U-shaped architecture. The core of our model relies on a crucial process of feature fusion, which seamlessly merges transformer-based spatial representations with context-aware representations.

\par In our approach's initial phase, the Swin Transformer and the UNet encoder process the input image simultaneously. To capture global and local information effectively, we combine the features generated by these two components at different scales using element-wise sum operations. These fused features are then fed as inputs for subsequent processing stages, allowing the integration of diverse information for improved analysis and detection.

\par Integrating the Swin Transformer and incorporating multiple stages of transformer layers and self-attention mechanisms into the Swin-Tempo framework allows the capture of meaningful spatial features, forming a vital foundation for capturing fine-grained details and global context from anatomical structures and lung nodule characteristics. Our experimental results demonstrate that this efficient fusion method significantly improves the performance of our network, achieving the best performance in lung nodule detection tasks.

%###################Integrating GRU with UNet's decoder###################
\subsection{Integrating GRU with UNet's Decoder}
\par Our approach involves the fusion of feature maps extracted from the contracting path of the Swin-Tempo model. Subsequently, these resulting fused feature maps are transmitted to the UNet decoder through skip connections. In the decoder section, our methodology adopts a conventional design commonly encountered in the UNet architecture, comprising five convolutional layers accompanied by up-sampling layers. 
\par Our proposed method for lung nodule detection views each CT scan as a video sequence and leverages GRU for sequential modeling. The application of GRU for sequential modeling tasks, such as analyzing CT scans, enables us to effectively capture temporal context and dynamic changes within the data. By integrating GRU into the 2D nodule detection framework, our model maintains an internal hidden state that proficiently processes sequential data and conveys temporal context from previously observed slices to subsequent ones. This integrated approach accommodates factors like respiratory motion and variations in nodule appearances across different slices, empowering the model to adapt to nodule characteristics across slices and yield more accurate and resilient detection outcomes.

\begin{figure*}[]
	\centering
	\includegraphics[height=0.33\textheight]{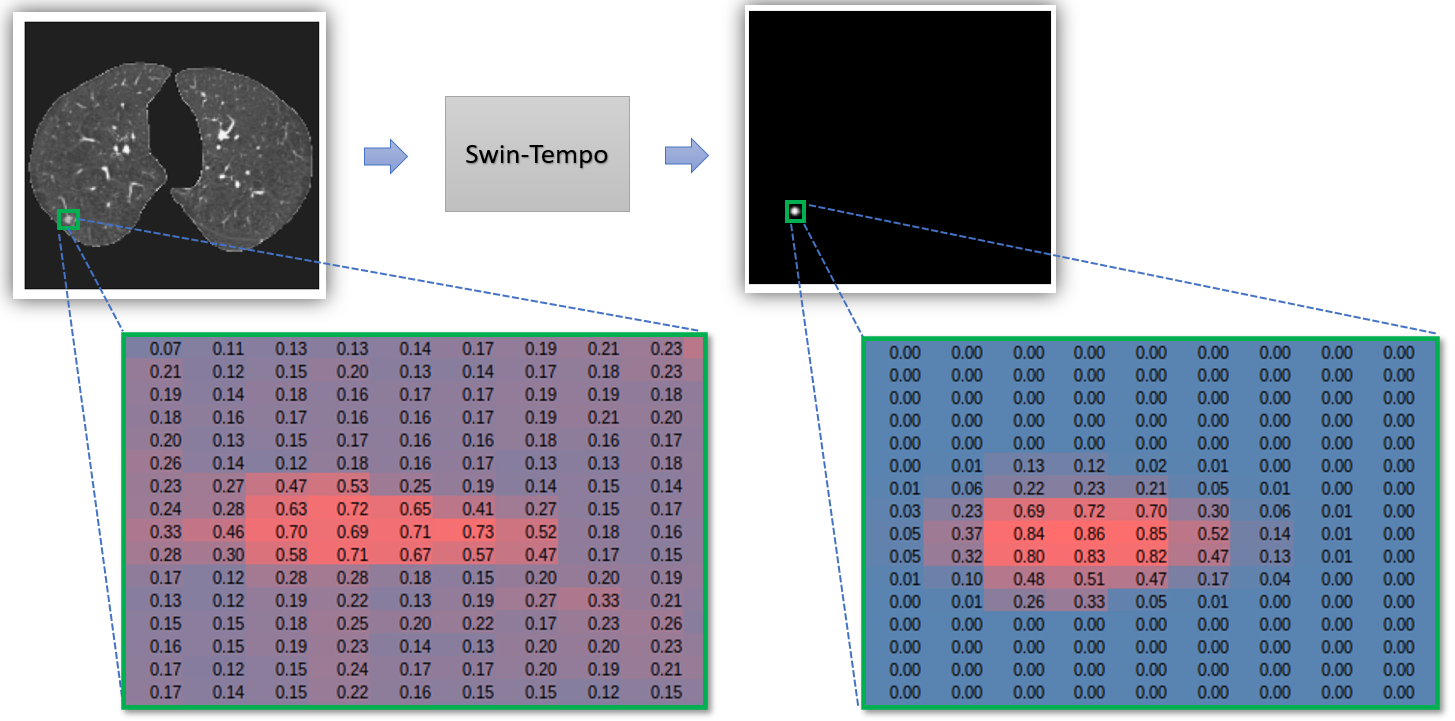}
	\caption{Swin-Tempo's output displays probability maps for lung nodules, which are subsequently converted into potential nodule masks through thresholding.}
	\label{seg11}
\end{figure*}

\subsection{Contour Finding to Detect Nodules from Segmentation Masks}

\par As previously mentioned, our proposed method involves sequentially processing CT scan slices to generate lung nodule segmentations that maintain the same 2D shape as the input image. These segmentations contain pixel values that serve as probabilistic indicators, reflecting the likelihood of a specific pixel corresponding to a lung nodule, as shown in Figure \ref{seg11}. We apply a thresholding operation to these probabilistic maps to extract potential nodule binary masks. Pixels exceeding the threshold indicate a significant likelihood of a nodule's presence, while pixels falling below the threshold suggest a lower likelihood.

\par The transformation of segmentation masks generated by the model into lung nodule detections follows a systematic procedure consisting of several computational steps. We utilize OpenCV's contour-finding function to identify each potential nodule's mask as a contour, enabling precise nodule detection. Further refinement is achieved by employing the DBScan clustering algorithm, which combines the 2D segmentations of nodule candidates obtained from individual slices to create a unified 3D representation of lung nodule detections.

\par Each nodule candidate is assigned a probability score determined by the maximum probability within its contour, and its 3D coordinates are represented in spherical coordinates (z, x, y, r), where each sphere signifies a potential nodule candidate. This probability score is a reliable indicator of the likelihood that the candidate genuinely represents a lung nodule.

\begin{figure*}[]
	\centering     
	\begin{tabular}{c}

			\label{cropping-orig}%
			\includegraphics[width=0.8\textwidth]{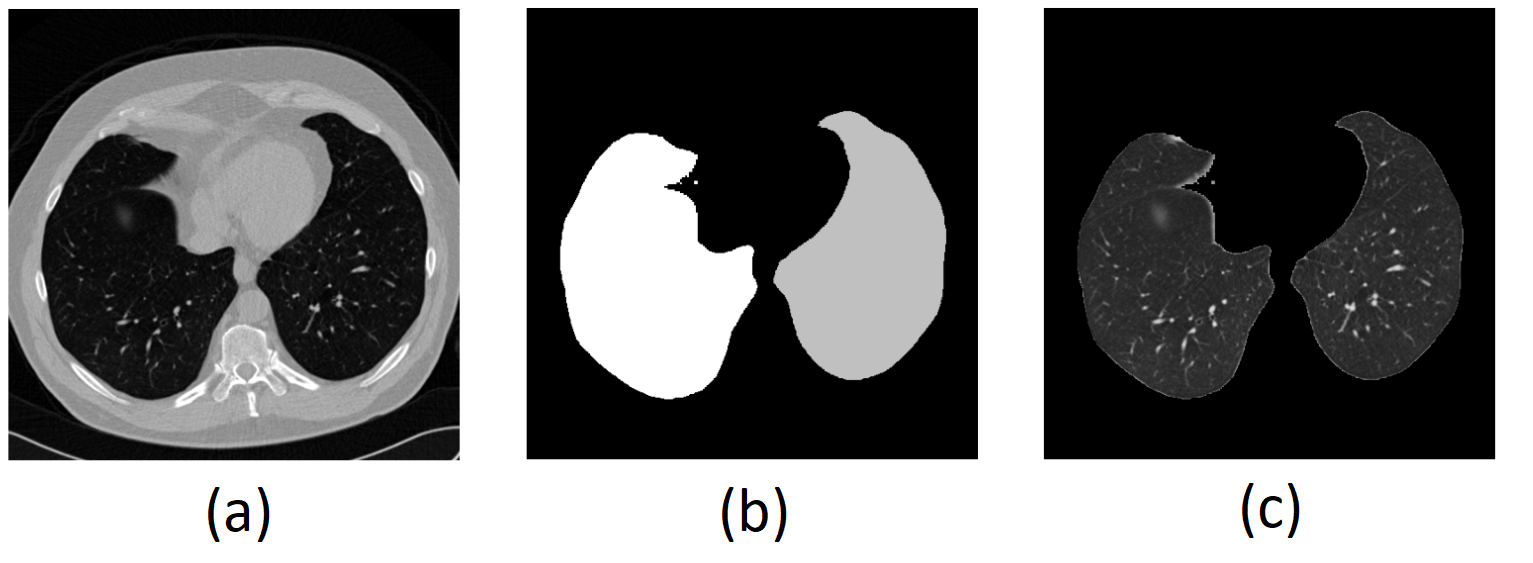}%
		
	\end{tabular}
	\caption{The example of preprocessing of a single slice.}
	\label{cropping}
\end{figure*}

\subsection{Loss Function}
Image segmentation, fundamentally a pixel classification task, employs a conventional loss function known as binary cross entropy \cite{yi2004automated}. The binary cross entropy loss function is defined as follows:

\begin{equation}
	\label{eq:binary_cross_entropy}
	\text{BCE}(y, \hat{y}) = -[y \cdot \log(\hat{y}) + (1 - y) \cdot \log(1 - \hat{y})]
\end{equation}

where  $y_i$ indicates the pixel i with the binary value (label) and $\hat{y_i}$ signifies the predicted probability.

\section{Experimental Results}\label{sec3}

\subsection{Dataset}\label{subsec31}

The large-scale benchmark data set LUNA16 \cite{kuan2017deep} was used to verify the detection performance of the proposed method. The LUNA16 dataset was released as part of the LUNA16 challenge in 2016. It contains 888 CT scans selected from LIDC-IDRI \cite{armato2011lung}. Four experienced radiologists labeled the annotations of all nodules. There are 1186 true positive nodules, and each true positive nodule was marked ‘1’ by at least three radiologists. The lung nodules are often in different sizes and variable shapes.

\subsection{Data Preprocessing}\label{subsec32}
In our data preprocessing approach for model input, we follow the methodology detailed in \cite{mkindu20233d}. Initially, we clip raw CT scans by setting Hounsfield units within the range of -1200 to 600, effectively removing irrelevant information. Subsequently, we employ a ground-truth lung segmentation mask to isolate lung regions, ensuring that our model primarily processes lung-related data. Finally, we standardize the preprocessed images with a mean of 0 and variance of 1, facilitating neural network analysis. These steps enhance the quality and relevance of our input data for accurate CT scan analysis.

\subsection{Implementation Details}\label{subsec33}

The Swin-Tempo model is implemented using Python 3.10 and PyTorch 1.12.1. To ensure data diversity and prevent overfitting, we apply various data augmentations to all training cases, including scaling, brightness adjustment, rotation, shearing, and translation. The input image size is 224x224, without cropping CT images, while the patch size for the Swin Transformer is set to four. Training is performed on an NVIDIA GeForce GTX 1080Ti graphics processing unit. This method involves utilizing pre-trained Swin Transformer weights to initialize the Swin-Tempo architecture, benefiting from the prior training on ImageNet-1K data \cite{krizhevsky2012imagenet}. Throughout the training process, we utilized the widely used Adam optimizer with a weight decay of 1e-4 for model optimization through backpropagation. These measures contribute to the model's robustness and efficiency in lung nodule detection tasks.

\subsection{Evaluation Metrics}\label{subsec34}

The Swin-Tempo model underwent a rigorous evaluation process encompassing tenfold cross-validation using the LUNA16 dataset. The evaluation of the detection model was centered on the performance metrics outlined by the LUNA16 challenge, with a primary emphasis on using free-response receiver operating characteristics (FROC) analysis at predefined false-positive rates per scan. These rates were set at 1/8, 1/4, 1/2, 1, 2, 4, and 8. The model's performance was then assessed by averaging these seven sensitivities within the context of CPM \cite{liu2017multiview}. Sensitivity was determined using Equation \ref{eq:sensitivity}, where TP represents true positives and FN represents false negatives.

\begin{equation}
	\label{eq:sensitivity}
	\text{Sensitivity} = \frac{\text{TP}}{\text{TP} + \text{FN}}
\end{equation}

\subsection{Ablation Study}

\begin{figure}[t]
	\centering
	\includegraphics[height=0.3\textheight]{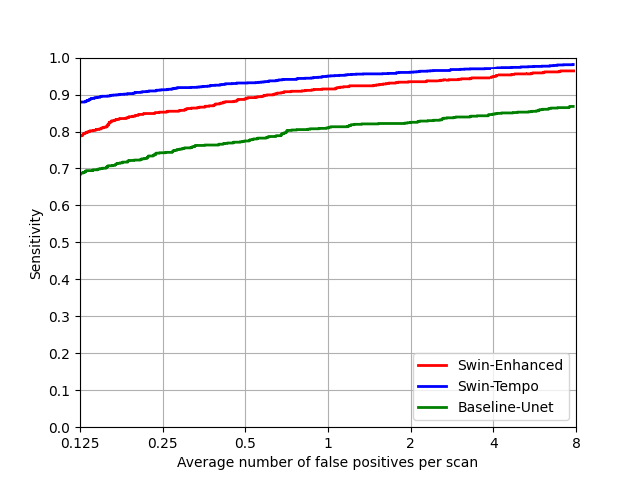}
	\caption{The comparative analysis of FROC curves among the models in the ablation study.}
	\label{froc}
\end{figure}

In this comprehensive ablation study, we systematically assess the performance and efficacy of our proposed model for lung nodule detection in CT scans. Our study unfolds in three distinct phases, each building upon the previous, as follows:

Firstly, we establish a baseline performance by training a conventional UNet model ("Baseline UNet") for lung nodule detection. Subsequently, in our second experiment, we enhance the UNet model by incorporating the Swin Transformer into the encoder architecture ("Swin-Enhanced"). This integration of the Swin Transformer operates in parallel with the UNet encoder, effectively bolstering feature extraction capabilities. In the final experimental phase, we extend the Swin-Enhanced model to formulate the "Swin-Tempo" model. Here, in addition to the Swin Transformer within the encoder, we introduce a decoder module based on GRU, thus enabling temporal awareness.

The performance evaluation of our ablation study is graphically presented in Figure \ref{froc}, offering a clear illustration of the Swin-Tempo model's superiority over its predecessors. These results underscore the effectiveness of our proposed Swin-Tempo model, which excels in achieving heightened sensitivity while concurrently minimizing false positives. This robust performance positions the Swin-Tempo model as a promising choice for precise and reliable lung nodule detection in medical imaging.

\begin{figure}[b]
	\centering
	\includegraphics[height=0.3\textheight]{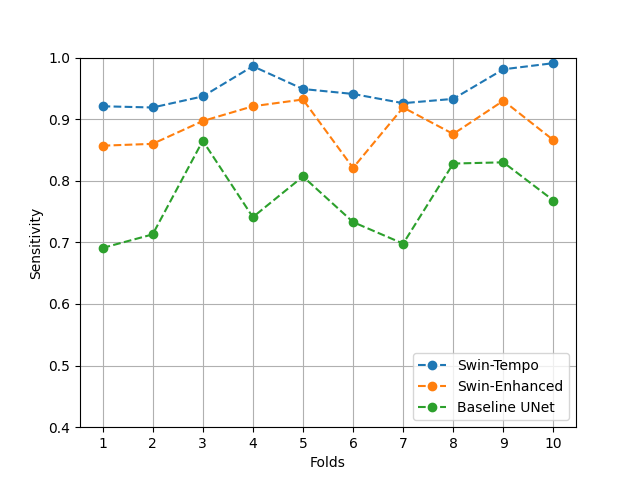}
	\caption{Sensitivity measure for different subsets of LUNA16 dataset. }
	\label{fold_wise}
\end{figure}

Additionally, the impact assessment of our proposed CAD scheme is conducted through ablation studies employing the LUNA16 dataset with a rigorous 10-fold cross-validation methodology, as represented in Figure \ref{fold_wise}. These experiments consistently reveal that the Swin-Enhanced model outperforms the Baseline UNet across all ten subsets during cross-validation. Furthermore, the superior performance of the Swin-Tempo model in comparison to the Swin-Enhanced model underscores the value of incorporating temporal awareness through the GRU-based decoder, which significantly enhances lung nodule detection performance. These discernible performance trends affirm the effectiveness and superiority of our proposed CAD scheme, elevating the overall performance of the lung nodule detection system.

Table \ref{tab13} displays the outcomes of the ablation models. The Baseline UNet model successfully identified 852 out of 1186 nodules in the LUNA16 dataset, representing 4484 candidates. By integrating the Swin Transformer into the Baseline UNet, the average sensitivity significantly improved from $78.14\%$ to $91.09\%$, demonstrating the advantage of employing the Swin Transformer for extracting global features and enhancing nodule detection. However, during our model evaluation, we observed that utilizing inter-slice information could decrease false positive rates. Consequently, incorporating the GRU into the Swin-Enhanced model resulted in an additional $6.75\%$ increase in average sensitivity, enabling the detection of 1125 nodules and representing 65 more candidates than the Swin-Enhanced model. These findings highlight the effectiveness of the GRU-based decoder in capturing temporal context and enhancing overall performance in lung nodule detection.

\begin{table*}[]
	\centering
	
	\caption{Comparison of the ablation model experiments.}
	\label{tab13}
	\begin{tabular}{cccc}
		\hline
		\textbf{Models} & \textbf{Candidates}  & \textbf{Detected nodules} &\textbf{Sensitivity}(\%)  \\
		
		\hline
		
		Baseline UNet   & $4484$  & $852$  & $78.14$  
		\\
		Swin-Enhanced   & $4613$  & $1081$   & $91.09$
		\\	
		Swin-Tempo      & $4678$  & $1125$   & $97.84$ 
		\\ 
		\hline
	\end{tabular}
\end{table*}

\subsection{Results Visualization}\label{subsec35}

\begin{figure*}[]
	\centering
	\includegraphics[height=0.73\textheight]{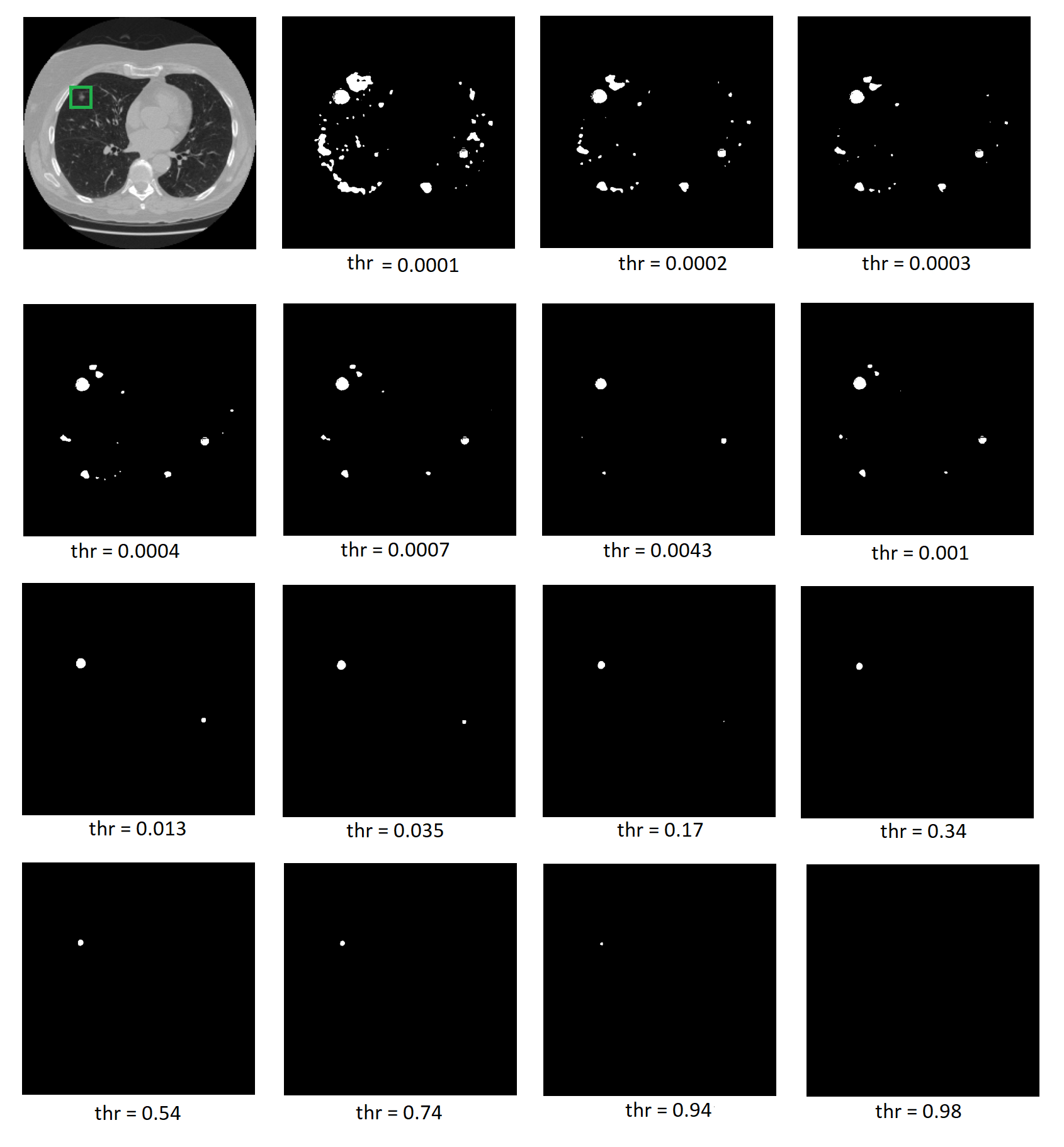}
	\caption{Swin-Tempo output for different threshold values.}
	\label{seg}
\end{figure*}

\begin{figure*}[]
	\centering
	\includegraphics[height=0.93\textheight]{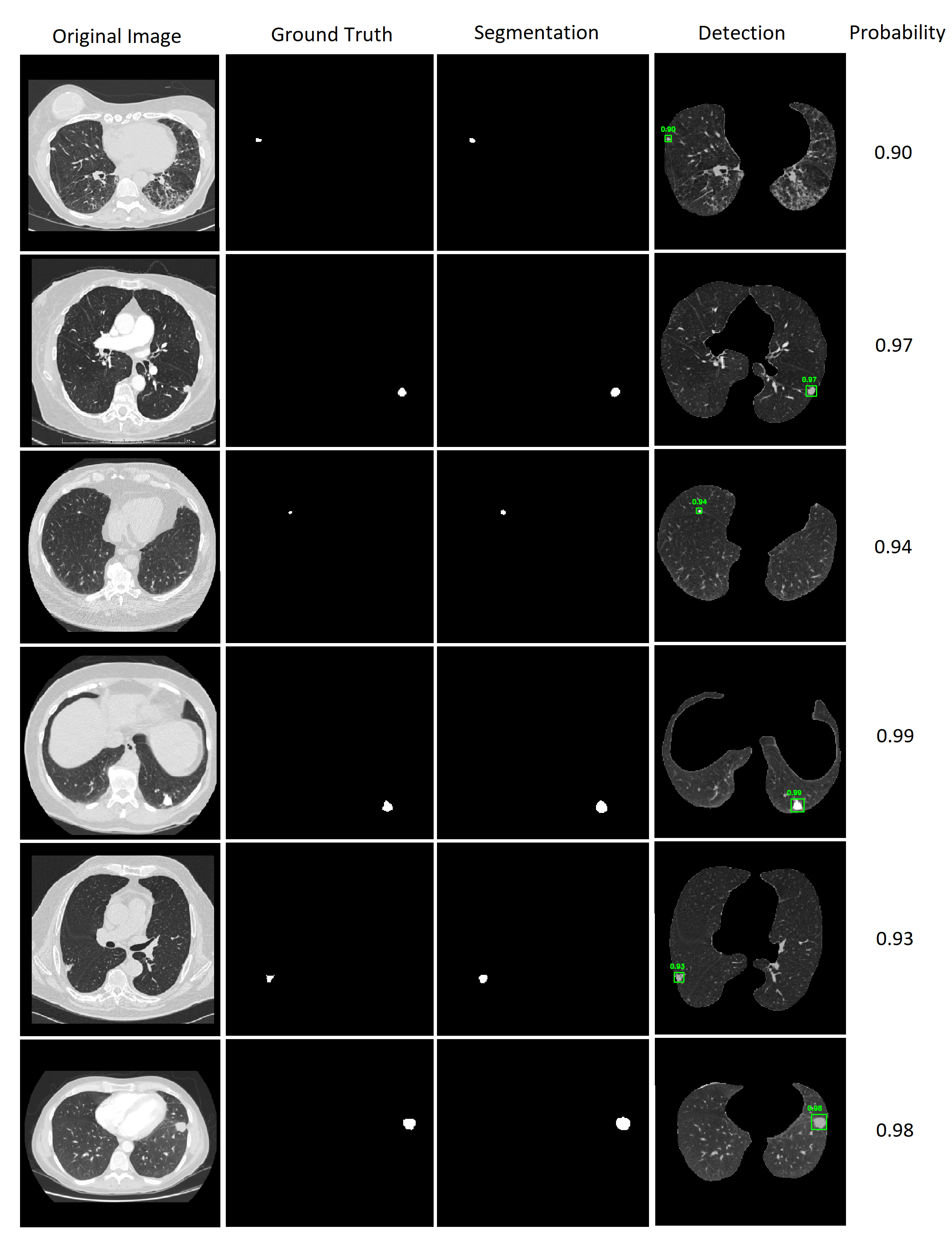}
	\caption{Swin-Tempo model's nodule detection results showcasing raw CT scans from the LUNA16 dataset, corresponding ground truth annotations derived from LIDC/IDRI, predicted nodule segmentations, and nodule probability scores.}
	\label{seg1}
\end{figure*}

Figure \ref{seg1} presents the outcomes of our novel lung nodule detection method across a range of scenarios. These scenarios were thoughtfully selected to encompass complex situations involving nodules of varying sizes, shapes, and locations within different lung regions. In this figure, each row represents a distinct case, starting with the original CT scan data in the first column. The second column displays ground truth annotations from the LUNA16 dataset, obtained from the LIDC/IDRI dataset. In the third column, we show the results of the Swin-Tempo model's 2D nodule segmentations, and later, these 2D segmentations are combined to form 3D segmentations. We further represent 3D nodule candidates using DBScan clustering. The fourth column illustrates the output of preprocessing steps, extracting the lung region from the background of the original image. The final column provides scores indicating the likelihood of a region containing a nodule based on the probability values derived from our model. Remarkably, even in these challenging cases, our model consistently delivers highly accurate predictions, demonstrating the effectiveness and robustness of our proposed methodologies.
We also include an illustrative example of the Swin-Tempo model's segmentation with various threshold values in Figure \ref{seg}, which will be incorporated into our model.

%%%%%%%%%%%%%%%%%%%%%%%%%%%%%%%%%%%%%%%%%%%%%%%%%%%%%%%%%%%%%%

%%%%%%%%%%%%%%%%%%%%%%%%%%%%%%%%%%%%%%%%%%%%%%%%%%%%%%%%%%%%%%

\begin{table*}[th]
	\caption{Performance of the models on the LUNA16 dataset.}\label{tab2}
	\begin{tabular*}{\textwidth}{@{\extracolsep\fill}lccD{.}{.}{2.2}D{.}{.}{2.2}}
		\toprule%
            \toprule
		Methods & Year & \multicolumn{2}{c}{Sensitivity(\%)} & \multicolumn{1}{c}{CPM(\%)} \\ \cmidrule{3-4}
		& &  \multicolumn{1}{c}{FPs/scan=1} & \multicolumn{1}{c}{FPs/scan=4} & \\ \midrule
		Zhang et al. \cite{zhang2023pulmonary} & 2021  & 91.8 & 94.8 & 89.3 \\
  
		Zhu et al. \cite{zhu2022channel} & 2022 & 91.7 & 95.2 & 89.5 \\ 
  
		Huang et al. \cite{huang2022one} & 2022 & 92.8 & 96.1 & 90.5 \\
		Mkindu et al. \cite{mkindu20233d} & 2023 & 93.8 & 96.6 & 91.1 \\
  
		Zhang et al. \cite{zhang2022attention} & 2022  & 86.4 & 91.8 & 91.2 \\
  
		Ozdemir et al. \cite{ozdemir20193d} & 2020  & 94.2 & 95.9 & 92.1 \\
  
            Cao et al. \cite{cao2020two} & 2020  & 93.6 & 95.7 & 92.5 \\   
            
            Zheng et al. \cite{zheng2021deep} & 2021  & 94.2 & 96.6 & 94.0 \\
            Chen et al. \cite{chen2023novel} & 2023  & 97.1 & 98.2 & 95.5 \\

		Our &   & 96.5 & 98.7 & 96.0 \\ \midrule \midrule
	\end{tabular*}
\end{table*}

\section{Discussion}

\par In this study, we proposed an innovative end-to-end Swin-Tempo framework for the automated detection of pulmonary nodules. This framework leverages a hybrid architecture, combining the Swin Transformer, RNN, and UNet models. By seamlessly integrating the Swin Transformer into the UNet architecture, our model effectively captures spatial context and harnesses the feature extraction capabilities of transformers. Furthermore, incorporating an RNN introduces dynamic network behavior, reducing computational overhead and enabling the model to process individual 2D slices while considering temporal information from preceding slices. Our experimental evaluation, conducted using the LUNA16 dataset and employing a ten-fold cross-validation methodology, demonstrates the superior performance of our model compared to state-of-the-art techniques for nodule detection on the same dataset.
As detailed in Table \ref{tab2}, our method achieved a high CPM value of $96.0\%$ and sensitivity score of $96.5\%$ and $98.7\%$ at 1 and 4 FPs/scan, respectively, while maintaining a low count of false-positive cases. These results highlight the reliability and efficiency of our approach in comparison with other traditional methods. It is worth noting that while two-stage methods may offer higher accuracy, they typically require more time and substantial GPU memory space due to their multi-step nature.

\par Cao et al. \cite{cao2020two} introduced a two-stage CNN-based structure that utilized a UNet segmentation model with a residual-dense structure in the first stage to identify the centroid of the mask, followed by a 3D CNN-based ensemble learning architecture in the second stage for false positive reduction. Chen et al. \cite{huang2022one} proposed a multi-view and multi-scale shared convolutional structure model for lung nodule segmentation and detection, achieving a CPM score of $95.5\%$, outperforming Cao's work. In our research, we have developed an alternative approach that surpasses their method regarding both sensitivity and the reduction of false-positive results. Our strategy involves using a simpler single-stage 2D UNet segmentation network with low complexity, yet it performs better. Furthermore, our model leverages the advantages of two distinct feature fusion techniques based on transformer-CNN architectures at different network depths, resulting in enhanced feature extraction and overall effectiveness.

\par The multi-scale model proposed by Mkindu et al. \cite{mkindu20233d}, which incorporates a local-global transformer architecture. However, it relies on the memory-intensive 3D Swin Transformer without CNN frameworks, which poses storage challenges. Moreover, it uses multi-scale 3D patches in the embedding layer. However, it lacks hierarchy, limiting its ability to extract spatial information at varying scales and receptive fields, particularly for small lung nodule detection. In contrast, our model balances memory efficiency and performance with its encoder-decoder structure and skip connections between multi-level feature maps. This architecture inherently captures features at multiple scales, ranging from the global context with larger receptive fields in the encoder to finer details with smaller receptive fields in the decoder. This design ensures robust performance detecting lung nodules of various sizes and contexts while efficiently utilizing computational resources. The hierarchical representation further enhances spatial understanding, making our model particularly effective in sensitive detection tasks, including identifying small nodules.

\par Zhu et al. \cite{zhu2022channel} incorporated attention mechanisms into a 3D convolutional neural network to improve feature extraction for predicting nodules. In terms of test outcomes, their CPM score fell short of ours. In contrast to the studies conducted in \cite{zhang2022attention, zhang2023pulmonary}, we utilized whole 2D image slices as input for our model without cropping and achieved high scores. This strategy allows our model to comprehensively capture structural information from the input image, facilitating a deep understanding of the lung region. Additionally, our approach emulates the capabilities of a 3D model by employing an RNN to extract spatial contextual features from 3D CT scans, all while maintaining a low number of parameters.

Generally, our proposed approach provides an effective solution for lung nodule detection, combining efficiency and accuracy. Our model achieves exceptional performance by utilizing 2D slices and an RNN-based design while addressing the computational and convergence challenges of 3D CNNs. These findings underscore the practicality and effectiveness of our approach in lung nodule detection, demonstrating the efficacy of the Swin-Tempo architecture for nodule detection tasks.

\section{Conclusion}\label{sec5}
In this study, we leverage a hybrid architecture by applying a 2D segmentation model to represent 3D lung nodules. Using video-like analysis, we extract spatial contextual information from CT scans without increasing computational demands. Combined with RNN, transformer and CNN architectures enhance nodule detection tasks. Extensive evaluation and rigorous experimentation consistently demonstrate the superiority of our proposed approach, surpassing existing methods.
Since medical image analysis is becoming increasingly computationally intensive, our model stands out for its exceptional performance, with significantly reduced resources. Our model is practical and accessible even on less powerful devices. Specifically, it can detect nodules in CT scans, particularly in environments lacking computational power.
Our model will be further enhanced by reducing false positive rates and improving diagnostic accuracy in the future. This will involve sequentially incorporating information from preceding and subsequent slices and continuous temporal context into the analysis. This innovative approach can detect lung nodules earlier, improving lung cancer diagnosis and detection.

\section*{Declarations}

\noindent
\bmhead{Funding} The authors declare that no funds, grants, or other support were received during the preparation of this manuscript.

\noindent
\textbf{Competing Interests} We declare that we have no known competing interests that could have appeared to influence the work reported in this paper.

\noindent
\textbf{Author Contributions} HJ: Methodology, Conceptualization, and writing—original draft preparation. KF: Supervision, Editing, Modification for the final layout. HA: Supervision, Review \& Editing.

\FloatBarrier
\newpage
\newpage

\bibliography{Manuscript}
\end{document}